# What are the Magic Clusters Observed in Graphene Chemical Vapor Deposition (CVD) Growth?

Junfeng Gao, and Feng Ding[1,2]*

Beijing Computational Science Research Center, Beijing, 100084, China.

Institute of Textiles and Clothing, Hong Kong Polytechnic University, Hong Kong, China.

To improve atomically controlled chemical vapor deposition (CVD) growth of graphene, understanding the evolution from various carbon species to a graphene nuclei on various catalyst surfaces is essential. Experimentally, an ultra-stable carbon cluster on Ru(0001), Rh(111) surfaces was observed, while its structure and formation process were still under highly debate. Using *ab initio* calculations and kinetic analyses, we disclosed a specific type of carbon clusters, composed of a $C_{21}$ core and a few dangling C atoms around, were exceptional stable in the size range from 21 to 27. The most stable one of them, an isomer of $C_{24}$ characterized as three dangling C atoms attached to the $C_{21}$ (denoted as $C_{21}$-3C), is the most promising candidate for the experimental observation. The ultra-stability of $C_{21}$-3C originates from both the stable core and the appropriate passivation of dangling carbon atoms by the catalyst surface.

Due to the recorded electron/hole mobility,[1] extremely high thermal conductivity[2, 3] and superior mechanical performance[4, 5], the two-dimensional graphene was broadly studied for many applications, such as ultrafast transistor,[6, 7] transparent and flexible electrode,[8] etc.[9]

Among the known methods of graphene synthesis, the chemical vapor deposition (CVD) is the best way to achieve high-quality single or few graphene layers with macroscopic area (up to 1000 inch$^2$) at a reasonable price, and therefore it was extensively explored both experimentally and theoretically.[10-28] In a CVD experiment, the transition metal surface, on which graphene is formed by the self-assembly of the dissociated or deposited C atoms or radicals, plays a crucial rule. To date, a cornucopia of transition metals, such as Ru,[11, 29] Rh,[30] Ir,[11] Cu,[17, 21] Ni,[31] Pt[23], Au[32] and their metal alloys (i. e. Ni/Mo[19] and Ni/Cu[20]) were proved to be suitable for CVD of graphene.

The first stage in the CVD growth of graphene—the nucleation of graphene on the catalyst surface determines both the density of nuclei and the concentration of grain boundaries in the final product. Therefore nucleation of graphene has been intensively studied recently.[10-18] Theoretically, the evolution of C clusters and the size-dependent C formations on the catalyst surface have been investigated to determine the nucleation barrier and size of nuclei under various growth conditions.[13] Previous theoretical explorations have shown that, by carefully optimizing the experimental condition, the size of nuclei can be controlled between N=20–100.[13-15] Therefore the structures and stabilities of carbon clusters in this range are crucial for understanding the graphene nucleation.

For graphene nucleation on Ru(0001) and Rh(111) surfaces, an astonishing experimental observation is the existence of ultra-stable carbon clusters of ~1 nm in diameter.[33-35] Although different feedstock (e.g., $C_2H_4$,[33, 34] $C_{12}H_{24}$[33] and $C_{60}$[35]) were used in the graphene synthesis, nearly all the decomposed C atoms on the catalyst terrace were observed in the form of the uniformed ultra-stable clusters at the temperature range of 870 K-1000 K, indicating that this cluster is highly superior to others in a very large size range. Mimicking the terminology of cluster science, we hereafter call them "magic clusters".

However, due to the high curvature near edge of the cluster and the influence of metal surface, scanning tunneling microscope (STM) was unable to disclose the precise atomic structure of the magic cluster. As its size is close to that of a coronene molecule ($C_{24}H_{12}$), the magic carbon cluster was once suspected to be the dehydrogenated coronene, i.e. the $C_{24}$ with seven 6-membered-rings, namely 7-6MRs[36] or 7-C6.[33, 34] Latterly, by systemically searching the sp2 hybridized C clusters on various metal surfaces, we suggested that a closed core-shell isomer of $C_{21}$, which was characterized as one hexagon surrounded by three hexagons and three pentagons alternatively, was much more stable than the 7-6MRs. Therefore, the $C_{21}$ isomer was highly considered as the observed magic cluster.[36]

In this manuscript, via density functional theory (DFT) calculations, we presented a new ultra-stable isomer of $C_{24}$, which has the core of the aforementioned $C_{21}$ and three dangling C atoms attached to the three pentagons (denoted as $C_{21}$-3C). The $C_{21}$-3C, with much lower formation energy than the $C_{21}$ and 7-6MRs, corresponds the global minimum and has the dominate population in a large cluster size range, 20≤N≤30 on the Ru(0001), Rh(111) surfaces. This study proposed a more probable candidate for the experimentally observed magic cluster in graphene CVD growth and explained its significant stability on both Ru(0001) and Rh(111) surfaces.

To demonstrate the stability and formation of the $C_{21}$-3C isomer, we proposed an isomerization from the 7-6MRs to $C_{21}$-3C on the Ru(0001) surface. Details of DFT calculations are in Supporting information 2(S2). Figure 1a shows the transformation from the 7-6MRs to the $C_{21}$-3C via the successive rearrangement of the three edge C-C bonds. The first bond rotation leads to a meta-stable structure with 1.32 eV higher in energy. After that, the second bond rotation becomes −0.08 eV lower than 7-6MRs in energy, implying this structure is thermodynamically more stable than the 7-6MRs. With the third bond rotation, the energy of the final $C_{21}$-3C is greatly reduced to −1.99 eV eventually. Such a remarkable energy reduction implies that the $C_{21}$-3C isomer is much more stable than the initial 7-6MRs. With these in hand, the population ratio of 7-6MRs to $C_{21}$-3C in thermal equilibrium can be estimated by

$$P_{7\text{-}6MR}/P_{C21\text{-}3C} = exp(-1.99\ eV/k_BT) = 10^{-11} \quad (1)$$

at the experimental temperature of 900 K, where $k_B$ is the Boltzmann constant and $T$ is the temperature. Besides, isomerization from the 7-6MRs to $C_{21}$-3C should be easily occurred at the experimental temperature of ~ 900 K as the energy barrier for a C-C bond rotation is only 1.38 eV (see Figure S3a), which slightly lower than the bond rotation of graphene zigzag edge on Co surfaces.[22] These results imply the observation of the 7-6MRs via dehydrogenated coronene on Ru(0001) surface is nearly impossible.[33]

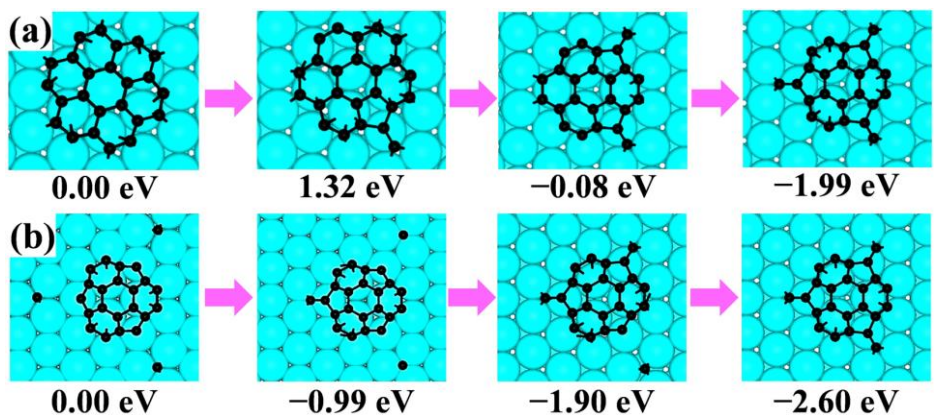

**Figure 1.** a) The isomerization from the 7-6MR to the $C_{21}$-3C of $C_{24}$ on Ru(0001) surface; b) a core-shelled $C_{21}$ adsorbs three carbon adatoms and becomes the $C_{21}$-3C on the Ru(0001) surface.

Next, let's consider the stability of the $C_{21}$-3C in relative to the previous proposed $C_{21}$ on the Ru(0001) surface.[36] A $C_{21}$-3C can be formed by a $C_{21}$ adsorbing three more C atoms from the catalyst surface (Figure 1b). The barriers of these atomic adsorptions are extremely low, less than 1.0 eV (see Figure S3b), and energy reductions are −0.99 eV, −1.90 eV and −2.6 eV, respectively. Considering the formation energy of a C monomer on the catalyst surface is about 0.55 eV higher than graphene, the formation energy of the $C_{21}$-3C is ~ 1.0 eV lower than the core-shelled $C_{21}$, the most stable cluster shown in the previous study.[36] Thereby, if there are sufficient C monomers on the catalyst surface, up to *[1−exp(−2.6 eV/$k_BT$)]* = ~99.99% of the core-shelled $C_{21}$ will be quickly transformed into the branched $C_{21}$-3C in the temperature range of 870 K−1000 K easily.

Above analysis clearly demonstrated the superior of branched $C_{21}$-3C over the dehydrogenated 7-6MRs and the core-shelled $C_{21}$. In order to understand the ultra-stability of the $C_{21}$-3C on the catalyst surface, let's exam its structural and electronic properties.

Both the $C_{21}$ and the $C_{21}$-3C have 9 active edge atoms. There is one dangling bond on each edge atom of the hexagon and two dangling bonds on each of the pentagon (for $C_{21}$) or the dangling carbon atom (for $C_{21}$-3C), thereby both $C_{21}$ and $C_{21}$-3C have 12 dangling bonds (see Figure 2a, d). However, the arrangement of paired dangling bonds on the edge of pentagon of $C_{21}$ is different from that on the dangling C atom of $C_{21}$-3C. On the edge of a pentagon, the C atom tends to be *sp³* hybridized and the four *σ* bonds tends to be tetrahedrally distributed in space. As a consequence, one dangling bond on the edge of a pentagon of $C_{21}$ flips upwards and thus was not efficiently passivated by the metal surface as evidenced by the differential charge densities near the edge atom (the red dangling bonds of Figure 2b-c) and

Figure S4). For only one dangling bond can be passivated by the metal surface, the edge C atom of the pentagon locates on the top of a metal atom.

Different from $C_{21}$, a dangling atom of $C_{21}$-3C is $sp^2$ hybridized with one double bond and two dangling bonds according to the analysis of differential charge densities (Figure 2e). The arrangement of the two dangling bonds tends to be in the same plane of the tilted pentagon as the characteristic of the $sp^2$ C network (see Figure 2d). As a consequence, all edge dangling bonds can be effectively passivated by the metal surface. To passivate both dangling bonds efficiently, all the three dangling atoms of the $C_{21}$-3C locate near hollow and bridge sites of the metal surface as can be clearly seen in Figure 2e, 2i and Figure S4. The length of outer-most C-C bond of $C_{21}$-3C is about 1.44 Å.

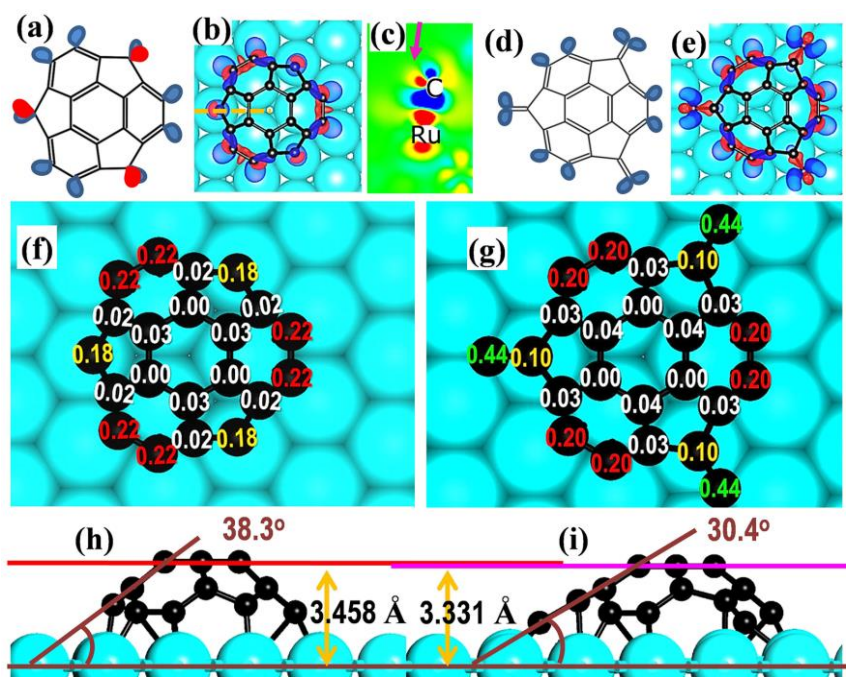

*Figure 2.* The schematics of dangling bonds of a) $C_{21}$ and d) $C_{21}$-3C; the top view of differential charge densities of b) $C_{21}$ and e) $C_{21}$-3C on Ru(0001) surface; c) the cross section view of the differential charge densities of the $C_{21}$@Ru(0001), the arrow points to the unpassivated dangling bond of the edge C atom; Bader charge analyses (in |e|) of f) $C_{21}$ and g) $C_{21}$-3C on Ru(0001) surface; h-i) the optimized $C_{21}$ and $C_{21}$-3C structures on the Ru(0001) surface.

Above analyses explicitly explained the high stability of $C_{21}$-3C in relative to the $C_{21}$ was stemmed from the more effective passivation of dangling atoms on the catalyst surface. As calculated in previous study,[36] the binding energies of $C_{21}$ on Ru(0001), Rh(111), Ir(111) surfaces were 16.9 eV, 18.1 eV and 17.6 eV, respectively, which were significantly larger than the core-shelled $C_{20}$ and 7-6MRs. Analyses with same method showed that the binding energies of $C_{21}$-3C were further increased to 22.5 eV, 23.1 eV and 21.9 eV on these metal surfaces, respectively, which were 5.6 eV, 7.0 eV and 4.3 eV higher than those of $C_{21}$ (Table S6), although both configurations possessed the same number of active edge atoms. The very large binding energies further validated the efficient passivation of the three dangling atoms of the $C_{21}$-3C.

The stronger binding between the $C_{21}$-3C and catalyst surface than $C_{21}$ is also verified from the Bader charge transfer analysis. As shown in Figure 2f-g, one can see the total charge transfer from the metal to the $C_{21}$ is about 2 |e|, which is much smaller than that from the metal to $C_{21}$-3C, ~ 3 |e|. The difference is from the edge of the pentagons, where the charge transferred to an edge atom of a pentagon of $C_{21}$ is about 0.18 |e| while transferred to a dangling C atom of C21-3C is about 0.44 |e|.

The height of the experimentally observed magic carbon cluster is ~1 Å higher than coronene on the catalyst surface.[34] As shown in Figure 2h-i and Table S5, the height of $C_{21}$-3C is very close to that of $C_{21}$, which was proved about 1.23 Å higher than that of coronene.[36] That is the height of $C_{21}$-3C is about 1.0~1.1 Å higher than coronene on catalyst surface, agreeing well with the experimental observation.[34]

Now, let's consider the stability of the $C_{21}$-3C on metal surface from the point view of aromaticity. There are two typical nonplanar π-conjugated polycyclic aromatic hydrocarbon (PAH) molecules of the similar size, $C_{20}H_{10}$ (corannulene, synthesized in 1991[37]) and $C_{21}H_{12}$ (sumanene, synthesized in 2003[38]), and both were extensively studied before.[39-43] As the $C_{5v}$ symmetry of $C_{20}$ is different from that of the Ru(0001) surface, $C_{20}$ on the Ru(0001) surface is less stable than $C_{21}$. In comparison with the $C_{21}$, the $C_{21}$-3C on Ru(0001) have very similar bonding situation with sumanene (see Figure S6 and Table S6 for detailed discussion). They both have very similar bond angel of a pentagon (~103°) and the bond lengths difference in the hexagons in a $C_{21}$-3C on Ru(0001) is only 0.014 Å, which is even smaller than that of sumanene (0.046 Å).[44] This implies that $C_{21}$-3C on metal surface can be viewed as a typical nonplanar π-conjugated aromatic C cluster.[42, 44]. In contrast, the $C_{21}$ on Ru surface has much larger bong angle on its pentagon (~ 107°) and bond length difference in its hexagons. It's also worth to note that the $C_{21}$-3C can be viewed as derivative of the extended π-conjugated sumanene as shown in Figure S6e and Ref. 45.

Moreover, the simulated STM image of the $C_{21}$-3C cluster on Ru(0001) is also in good agreement with the experimental ones (see Figure S7).[33, 35] It is interesting that the simulated STM image of $C_{21}$ is also very close to the experimental ones.[36] This is attributed to the invisibility of the three dangling atoms due to their lower positions on the metal surface (Figure 2i). Same as that on Ru(0001) surface, the STM images of $C_{21}$ and $C_{21}$-3C on Rh(111) and Ir(111) surfaces (Figure S8) are very similar to each other.[34]

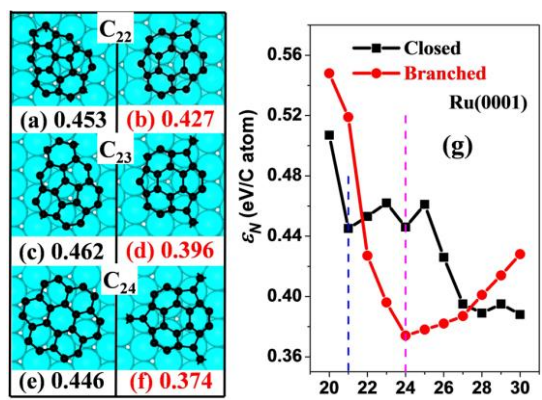

*Figure 3.* The most stable core-shell $sp^2$ network and branched $C_N$ (N = 22, 23, 24) clusters on Ru(0001) surface (left panel) and the formation energies $\varepsilon_N$ (in eV/C atom) of the most stable $C_N$ (N=20~30) clusters of each type on the Ru(0001) surface.

With above compelling evidences and analyses, the $C_{21}$-3C is expected to be the observed magic cluster in graphene CVD growth. As shown in previous study, the $C_{21}$ has very similar stability with the 7-6MR isomer on Ru(0001) and Rh(111) surfaces, and thus it corresponds a local minimum in the size range of N<24.[36] How is the stability of the $C_{21}$-3C in relative to $C_{21}$ and others? As shown in Figure 3, another two carbon clusters with one and two dangling edge atoms, namely $C_{21}$-1C (Figure 3b) and $C_{21}$-2C, respectively (Figure 3d), are also very stable. In order to fully address the stability of $C_{21}$-3C and its role of graphene nucleation, we systematically explored two serious of C clusters, (i) those are composed of pentagons and hexagons only as studied previously[13, 14, 18, 36] and (ii) those with one or a few dangling C atoms attached to the pentagons of the $C_{21}$ core, on three catalyst surfaces−Ru(0001), Rh(111) and Ir(111). The most stable structures and corresponding formation energies of the clusters on Ru(0001), Rh(111) and Ir(111) surface were shown in Figure 3g, S9, S10 and S11, respectively. Besides, in order to find the lowest-energy structures in this regime, many different isomers were also considered (for example, some explored structures of $C_{24}$, $C_{28}$, $C_{29}$ and $C_{30}$ are shown in the Figure S12).

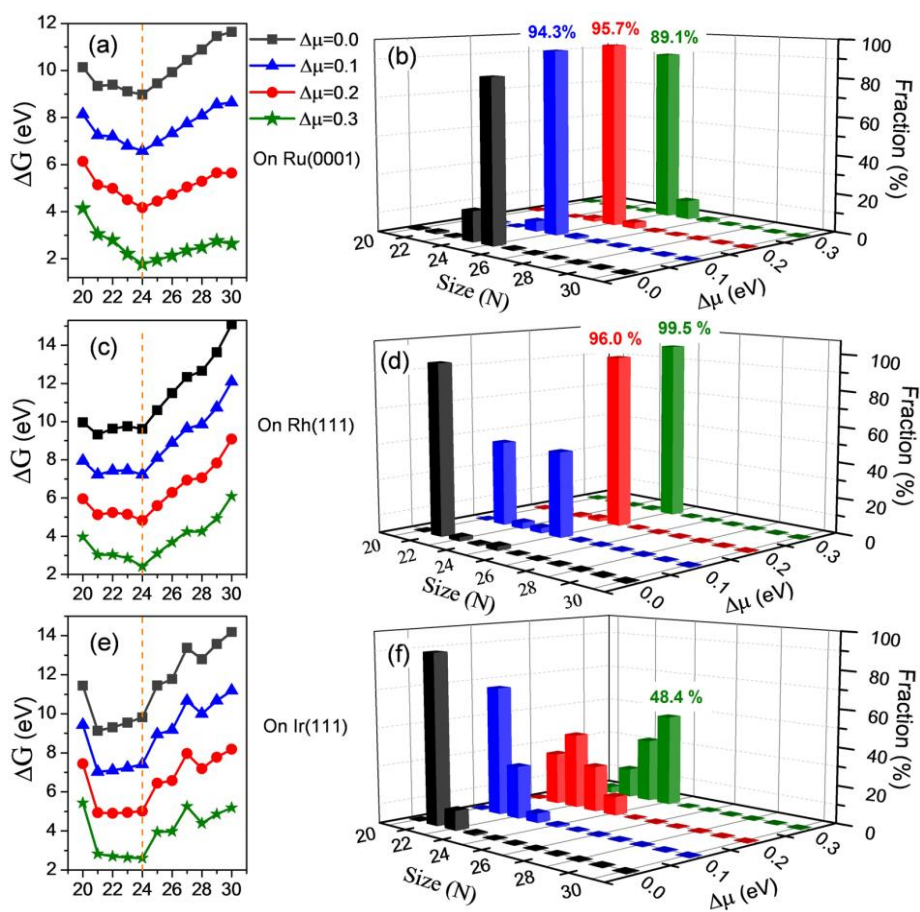

*Figure 4.* The Gibbs free energies and populations of $C_N$ (20≤N≤30) clusters under different condition of graphene growth (denoted by the chemical potential difference between feedstock and graphene, Δμ=0.0, 0.1, 0.2 and 0.3 eV) on a, b) Ru(0001); c, d) Rh(111), and e, f) Ir(111) surfaces, respectively.

For $C_N$ on all the three catalyst surfaces, in the explored cluster size region (20≤N≤30), the $C_{21}$-3C represents a global minimum (see Figure 3g). On Ru(0001) surface, the formation energy per C atom of C cluster keep dropping from N=20 to N=24 and then keep rising from N=24 to N=30. This characteristic makes $C_{21}$-3C a deep global minimum in the size range of 20≤N≤30. Although $C_{21}$ and 7-6MRs also correspond two local minima in the energy curve of the core-shell clusters (vertical dash lines in Figure 3g),[36] their formation energies are much higher than the branched C clusters on Ru(0001). This means that the branched C clusters possess the highest stability in the vicinity of size range of N=24 on Ru(0001) surface. It is important to note that the C clusters on both Rh(111) and Ir(111) surfaces exhibit very similar characteristics as that on Ru(0001) (Figure S9-11), implying that the above analyses may be universal for graphene CVD growth on metal surfaces.

Above analyses demonstrate that the branched $C_{21}$-3C isomer is superior in a large size region. To quantify the population of the $C_{21}$-3C cluster on metal surface under different growth condition, let's introduce the chemical potential difference between the C atom in feedstock and in graphene on the catalyst surface, $\Delta\mu$. According to the classical theory of crystal nucleation,[46] the population of a cluster can be estimated by

$$P_N = exp(-\Delta G/kT), \qquad (2)$$

where $\Delta G = E_N - N \times \Delta\mu$ is the Gibbs free energy and $E_N = N \times \varepsilon_N$ is the total formation energy of cluster $C_N$. The calculated $\Delta G$ and normalized fraction $P_N$ in the size range $20 \leq N \leq 30$ with $\Delta\mu = 0.0, 0.1, 0.2$ and $0.3$ eV are shown in Figure 4, respectively. The kinetic stability of the clusters on the three catalyst surfaces is presented as following:

(i) On the Ru(0001) surface, the Gibbs free energy of the $C_{21}$-3C cluster is a deep local minimum in a large range of $\Delta\mu$ (Figure 4a). Due to the exceptional stability, the population of the $C_{21}$-3C always dominate in the explored size range (>89%) (Figure 4b).

(ii) On the Rh(111) surface, the $C_{21}$ has the lower formation energy than $C_{21}$-3C at $\Delta\mu=0.0$ eV, but when $\Delta\mu$ becomes slightly larger, i.e., $\Delta\mu>0.2$ eV (common condition), the $C_{21}$-3C becomes more stable than $C_{21}$ (Figure 4c). Thus the dominating of $C_{21}$-3C can be only observed in the range of $\Delta\mu \geq 0.2$ eV (Figure 4d).

(iii) On the Ir(111) surface, the small clusters, $C_{21}$, branched $C_{22}$ and $C_{23}$ are more stable than the $C_{21}$-3C in the range of $\Delta\mu<0.2$ eV (Figure 4e). The dominating clusters is $C_{21}$ at $\Delta\mu=0.0$ eV but no dominating clusters until $\Delta\mu>0.3$ eV (Figure 4f).

Above analyses indicate that the probability of observing the magic cluster on the three metal surfaces follows the order of Ru(0001)>Rh(111)>Ir(111). Such a trend is in perfect agreement with the known experimental facts. Experimentally, most reported magic clusters were on the Ru(0001),[33, 35] and a few were on the Rh(111)[35]. Although Ir(111) was considered as a very similar catalyst surface to Rh(111) and Ru(0001) for graphene growth, there's no uniformed C clusters on it.[47]

In summary, using DFT calculations, we systemically investigated the stability of $C_N$ clusters ($20 \leq N \leq 30$) on Ru(0001), Rh(111) and Ir(111) surfaces. Our study revealed a new type of ultra-stable carbon clusters—branched $sp^2$ carbon networks. The most stable one of them, $C_{21}$-3C, was predicated to be the experimentally observed magic cluster in graphene CVD growth on both Ru(001) and Rh(111) surfaces by stability and population analyses. Besides, we also successfully explained why the uniformed magic sized cluster only can be seen on Ru(0001) and Rh(111) surfaces but not on Ir(111) surface. This study provided a deep insight into the nucleation of graphene in the CVD growth processes and the formation mechanism of magic sized clusters might be used for the synthesis of graphene quantum dots.

**Acknowledgements**

We thank Prof. Zhenyang Lin of HKUST for useful discussion and acknowledge the supported by National Natural Science Foundation of China (No. 11304008), Development Fund of China Academy of Engineering Physics (No. 2013B0302056) and China Postdoctoral Science Foundation (No. 2014T70030, 2013M530019). The work was carried out at National Supercomputer Center in Tianjin, and the calculations were performed on TianHe-1(A).